\documentclass[aps,prl,floatfix,twocolumn,superscriptaddress,amsmath,amssymb]{revtex4}

 \usepackage{graphicx}% Include figure files	
 \usepackage{amssymb,amsmath}
\usepackage{epstopdf}
\usepackage{hyperref} % hyperlinks package

\begin{document}
  \title{Evanescent field optical readout of graphene mechanical motion at room temperature}
  
 \author{Robin M. Cole}
 \author{George A. Brawley}
 \affiliation{Queensland Quantum Optics Laboratory, University of Queensland
Brisbane, QLD 4072, Australia.}
 
 \author{Vivek P. Adiga}
\affiliation{School of Applied and Engineering Physics, Cornell University, 205 Clark Hall, Ithaca, New York 14853, USA.}
 
\author{Roberto De Alba}
\author{Jeevak M. Parpia}
 \affiliation{Department of Physics, Cornell University, 109 Clark Hall, Ithaca, New York 14853, USA.}
 
 \author{Bojan Ilic}
 \affiliation{Cornell Nanoscale Science and Technology Facility, Cornell University, 250J Duffield Hall, Ithaca, New York 14853, USA.}
 
 \author{Harold G. Craighead}
\affiliation{School of Applied and Engineering Physics, Cornell University, 205 Clark Hall, Ithaca, New York 14853, USA.}
 
\author{Warwick P. Bowen}
%\email{wbowen@physics.uq.edu.au}
\affiliation{Queensland Quantum Optics Laboratory, University of Queensland
Brisbane, QLD 4072, Australia.}
 
\begin{abstract} % 250 words
Graphene mechanical resonators have recently attracted considerable attention for use in precision force and mass sensing applications. To date, readout of their oscillatory motion has typically required cryogenic conditions to achieve high sensitivity, restricting their range of applications. Here we report the first demonstration of evanescent optical readout of graphene motion, using a scheme which does not require cryogenic conditions and exhibits enhanced sensitivity and bandwidth at room temperature. We utilise a high $Q$ microsphere to enable evanescent readout of a 70 $\mu$m diameter graphene drum resonator with a signal-to-noise ratio of greater than 25 dB, corresponding to a transduction sensitivity of $S_{N}^{1/2} = $ 2.6 $\times 10^{-13}$ m $\mathrm{Hz}^{-1/2}$. The sensitivity of force measurements using this resonator is limited by the thermal noise driving the resonator, corresponding to a force sensitivity of $F_{min} = 1.5 \times 10^{-16}$ N ${\mathrm{Hz}}^{-1/2}$ with a bandwidth of 35 kHz at room temperature (T = 300 K). Measurements on a 30 $\mu$m graphene drum had sufficient sensitivity to resolve the lowest three thermally driven mechanical resonances.
\end{abstract}

\maketitle % article length 3500 words
% use calibri for figures

\begin{figure}[t]
  \begin{center}
  \includegraphics[width=\columnwidth]{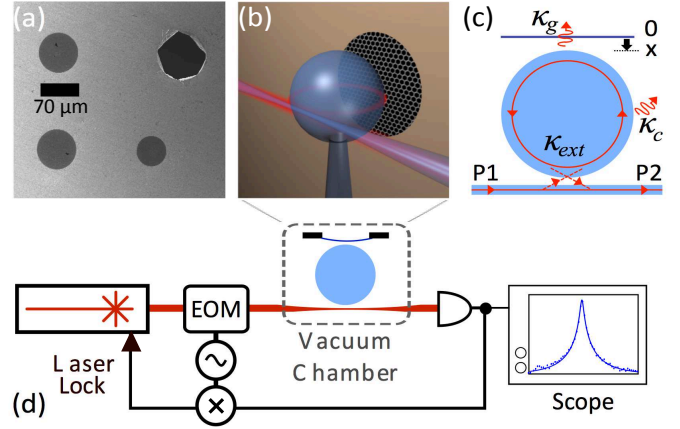}
\end{center}
\caption{\label{fig1} (a) SEM image of a graphene drum resonator's, a broken drum shows the contrast due to the graphene. (b) Illustration showing a microsphere with a graphene resonator in the evanescent field. (c) Illustration showing the optical coupling rates associated with the cavity measurement, discussed in the text. (d) Schematic of the measurement system.}
\end{figure}

Micro and nano-electro-mechanical (NEMS) force sensors are broadly applied in accelerometry\cite{Krause2012}, magnetometry\cite{Forstner2014}, thermometry\cite{Harris2013}, navigation\cite{Krishnan2012}, geodesy\cite{Yazdi1998}, medical diagnosis\cite{Maluf1995}, and have a range of specialised applications in areas such as atomic force microscopy\cite{SrinivasanNL2011}, nanoscale spin-resonance imaging \cite{Rugar2004}, and quantum information science\cite{LaHaye2009}. Currently silicon is the material of choice for fabricating precision NEMS force sensors, enabling the production of devices with both high sensitivity and large bandwidths. However graphene resonators have excellent mechanical properties and exceptionally low mass per unit area\cite{GeimNAT2007, BartonNL2011, Huttel2009}, which makes them attractive candidates for use in ultra-sensitive force\cite{EichlerNNan2011, BunchSCI2007, Moser2013, Stapfner2013} and mass\cite{ChasteNN2012, JensenNN2008, Chiu2008, Yang2006} measurements, as well as for quantum optomechanics\cite{SongARXIV2014, WeberNL2014, SinghARXIV2014}. 

In a generic NEMS force sensor, the motion of a compliant mechanical resonator is tracked optically or electrically as it responds to external forces\cite{Ekinci2005}. To reduce the measurement noise and enhance the measurement bandwidth a new approach has been developed where an optical or microwave cavity is used to control and readout the motion of the resonator\cite{Marquardt2009, Gavartin2012}. These cavity optomechanical systems allow force sensitivity at the fundamental thermomechanical noise floor $F_{min} = \sqrt{4m_{\mathrm{eff}}\Gamma_{m} k_{B}\mathrm{T}}$ in N ${\mathrm{Hz}}^{-1/2}$ where $m_{\mathrm{eff}}$ is the effective mass of the mechanical resonator, $\Gamma_{m}$ its mechanical dissipation rate, $k_{B}$ is the Boltzmann constant and T its temperature. Here we place graphene resonators into a cavity optomechanical system and demonstrate the sensitive readout of their motion at room temperatures, thus paving the way for the more widespread application of graphene NEMS force sensors.

Graphene's large electrical conductivity\cite{CastroNeto2009} and relatively straightforward integration into circuits have enabled sensitive motion readout using microwave electro-mechanical measurement systems\cite{BunchSCI2007}. However this approach requires cryogenic cooling of the electrical amplifiers owing to the relatively high Johnson (electrical) noise floor which severely limits the range of potential applications\cite{WeberNL2014, SinghARXIV2014}. Overcoming this limitation, measurement systems utilising visible optical wavelengths have been demonstrated at room temperatures \cite{Carr1997, BunchSCI2007, BartonNL2012}. The alternative approach described here places the graphene in the evanescent field of a high $Q$ silica microsphere, where the motion of the graphene is imparted onto the phase of the optical field circulating within the cavity. Crucially, in this arrangement the signal due to the graphene resonator is enhanced by the high optical $Q$ factor and the steep gradient of the evanescent field, lifting the signal above of the noise floor and enabling motion readout with large SNR\cite{Forstner2012a}. 

\begin{figure}[t]
  \begin{center}
  \includegraphics[width=\columnwidth]{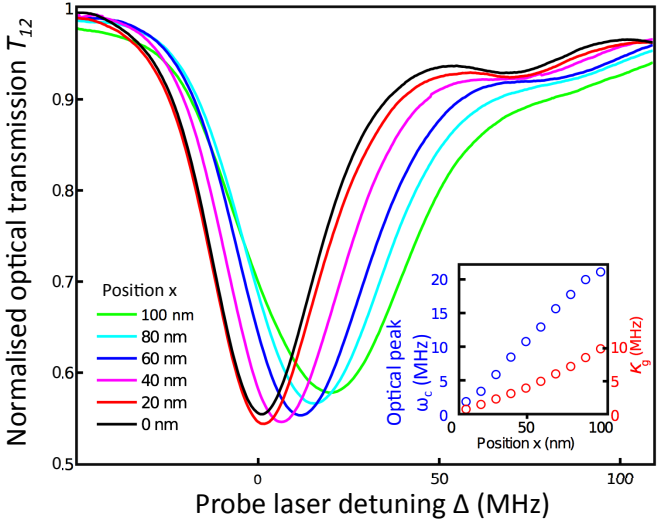}
\end{center}
\caption{\label{fig2} Experimentally measured normalised transmission ($T_{12}$) spectra with position of the graphene resonator $x$ within the evanescent field. At 0 nm the graphene is far from the microsphere surface, at 100 nm it is almost touching. Laser detuning is defined as $\Delta = 0$ at $\omega_{c}$ for the $x$= 0 nm spectra. (inset) Extracted optical resonance frequency $\omega_{c}$ and loss due to the graphene $\kappa_{g}$.}
\end{figure}

An SEM image showing graphene circular drum resonator's studied in this work is shown in Fig.\ref{fig1}(a). The graphene is produced by chemical vapour deposition (CVD) on a copper foil and transferred to a silicon substrate\cite{BartonNL2011} with pre-patterned circular through holes. Measurements have been performed on resonators with diameters in the range of d = 30 to 70 $\mu$m that support fundamental mechanical resonances in the range of $\omega_{m}/2\pi \approx$ 1 to 5 MHz\cite{Adiga2013, BartonNL2011}. An illustration of the motion readout measurement is shown in Fig.\ref{fig1}(b). A graphene resonator is shown placed within the evanescent field of a microsphere, typically at a distance of 50 - 100 nm from the microsphere surface. A tapered optical fibre is used to couple laser light to an optical resonance of the microsphere\cite{ChowOE2012}. Referring to the illustration in Fig.\ref{fig1}(c), we consider the effect that a small displacement $\partial x$ of the graphene resonator produces on the optical resonance. The displacement shifts the resonance by an amount $\partial\omega_{c}$, where the magnitude of this shift is determined by the optomechanical coupling strength $G$ that quantifies the overlap of the optical field with the graphene resonator, such that $G = \partial\omega_{c}/\partial x$. In addition to this dispersive shift, the displacement increases the optical linewidth (full width at half maximum) $\kappa = \kappa_{c} + \kappa_{g} + \kappa_{ext}$ where $\kappa_{c}$ is the internal cavity loss rate due to the intrinsic losses of the cavity (e.g. due to surface roughness), $\kappa_{ext}$ is the external coupling rate between the taper and optical resonance, and $\kappa_{g}$ is the optical loss due to the graphene (i.e. due to its 2.3$\%$ absorption). The sensitivity of motion readout measurements depends on the values of $G$ and $\kappa$, which depend on the particular mechanical and optical resonators under measurement. Here we show that despite the optical loss introduced by the graphene, motion readout measurements with high sensitivity are possible using cavity enhanced evanescent sensing.

The measurement system is shown in Fig.\ref{fig1}(d). To characterise $G$ and $\kappa$  a 780 nm diode laser outputting power P1 is swept over an optical resonance of the microsphere and the transmitted optical power P2 is recorded as a function of the position $x$ of the graphene within the evanescent field. Fig.\ref{fig2} shows spectra of the normalised optical transmission $T_{12}$ = P2/P1. These measurements were performed with a 70 $\mu$m diameter graphene resonator and a 60 $\mu$m diameter microsphere. Both $\omega_{c}$ and $\kappa$ are modified by the presence of the graphene. The spectra are described by\cite{MonifiJLT2012} $T_{12} = \frac{(\kappa_{o} + \kappa_{g} - \kappa_{ext})^{2} + 4\Delta^{2} } {(\kappa_{0} +\kappa_{g} + \kappa_{ext})^{2} + 4\Delta^{2}} $, where $\Delta = \omega_{p} - \omega_{c}$ is the detuning of the incident laser pump $\omega_{p}$ from the optical resonance $\omega_{c}$. A fit to a spectra recorded without the graphene present gave the intrinsic cavity loss rate $\kappa_{c}/2\pi$ = 10.2 MHz, corresponding to an optical quality factor $Q = \omega_{c}/\kappa_{c} = 3\times 10^{7}$. Fitting spectra to $\partial\omega_{c} = G \partial x$ yields the optomechanical coupling coefficient $G/2\pi$ = 0.21 MHz/nm. Since $\kappa_{c}$ is a fixed property of the optical resonance, increases in $\kappa$ are attributed to the optical loss introduced by the graphene $\kappa_{g}$. The extracted values of $\kappa_{g}$ with position $x$ are shown in the inset of Fig.\ref{fig2}. Extraction of $\kappa_{g}$ allows calculation of the optical power dissipated by the graphene, discussed later. 

To demonstrate the use of graphene resonators as sensitive force sensors we present experimental data showing the motion readout of the mean-square displacement $\langle x(t)^2 \rangle$ of the resonator arising from the random thermal (Brownian) force fluctuations $F_{\mathrm{T}}$ driving the resonator. These fluctuations are distributed in frequency $\omega$ according to $x(\omega) = F_{\mathrm{T}}\chi_{m}(\omega)$ where the resonator's Lorentzian-shaped susceptibility $\chi_{m}(\omega) = [m_{\mathrm{eff}}(\omega_{m}^{2} -\omega^{2} -i\Gamma_{m}\omega)]^{-1}$, and $F_{\mathrm{T}}$ is modelled as a white-noise. The power-spectral-density (PSD) of the random displacements of the resonator is given by $S_{xx}(\omega)= 4m_{\mathrm{eff}}\Gamma_{m}k_{B}\mathrm{T}|\chi_{m}^{2}(\omega)|$\cite{Ekinci2004}. The optomechanical coupling $G$ transduces the motion of the resonator into frequency noise on the optical resonance via $G = \sqrt{S_{\omega\omega}/S_{xx}}$. The measured signal is given by $S = S_{\omega\omega} + S_{N}$ where $S_{N}$ is the measurement noise (e.g. due to detector noise). At $\omega = \omega_{m}$ the measurement SNR $= \frac{S_{\omega\omega}}{ S_{N}} = \frac{4k_{B}\mathrm{T}}{m_{\mathrm{eff}} \Gamma_{m} \omega_{m}^{2} S_{N}}$, which on rearranging gives the noise limited measurement uncertainty

\begin{equation}\label{Sensitivity}
S_{N}^{1/2} = \left[\frac{4k_{B}\mathrm{T}}{m_{\mathrm{eff}}\Gamma_{m}\omega_{m}^{2}\mathrm{SNR}}\right]^{1/2} 
\end{equation}

\begin{figure}[t]
  \begin{center}
  \includegraphics[width=\columnwidth]{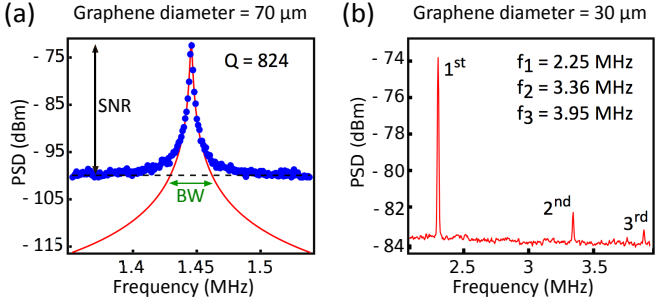}
\end{center}
\caption{\label{fig3} (a) Measured spectrum showing the fundamental mechanical resonance of a 70 $\mu m$ diameter graphene drum, spectrum analyser resolution bandwidth (RBW) = 100 Hz. (b)  Spectra showing the three lowest order resonances of a 30 $\mu m$ diameter resonator, RBW = 10 kHz.}
\end{figure}

Readout of graphene motion is performed using the system shown in Fig \ref{fig1}(d). The taper and graphene resonator are mounted on nano-positioning stages and aligned around the microsphere. To ensure high mechanical $Q$, the measurement is performed within a vacuum chamber that is maintained at a pressure of $3\times 10^{-7}$ Torr using a vibration free ion pump. The output from the laser is fibre coupled and passes through an electro-optic modulator (EOM) which applies a 100 MHz phase modulation to produce the error signal for the Pound-Drever-Hall (PDH) laser lock. The laser is locked to the side of the optical resonance with a fixed detuning of $\Delta = \kappa/2$, ensuring maximum transduction sensitivity. The $T_{12}$ transmitted light is incident on a low-noise photodetector, and monitored using an oscilloscope and a spectrum analyser. The measured mechanical spectra for a diameter d = 70 $\mu m$ resonator is shown in Fig.\ref{fig3}(a), measurement data marked with blue dots. The observed mechanical resonance is fit to a Lorentzian line-shape (red line) with the fit to the white-noise background indicated by the black dashed line. The mechanical resonance peak rises 25 dB above the flat measurement background, defining the measurement SNR. Assuming pristine monolayer graphene the effective mass of the fundamental drum mode of the resonator is calculated from\cite{WeberNL2014} $m_{\mathrm{eff}} = 0.27\pi r^{2}\rho_{2D}$ = 7.9 $\times 10^{-16}$kg where the mass density $\rho_{2D} =$ 7.9$\times 10^{-19}$ kg /$\mu m^{2}$ and r = d/2. From the Lorentzian fit, $\omega_{m}/2\pi$ = 1.44 MHz and $\Gamma_{m}/2\pi$ = 1.75 kHz. Substituting these values into Eq.\ref{Sensitivity} yields the transduction sensitivity of $S_{N}^{1/2} = $ 2.6 $\times 10^{-13}$ m $\mathrm{Hz}^{-1/2}$, which is currently limited by the $Q$ factor of the microsphere. As a force sensor this graphene resonator would enable measurements with a force sensitivity of $F_{min} = 1.5 \times 10^{-16}$ N ${\mathrm{Hz}}^{-1/2}$ at room temperature (T = 300 K). The usable bandwidth in this scenario is given by the frequency range for which the mechanical resonance is resolved with a SNR > 1, marked with a green arrow in Fig.\ref{fig3}(a). For this resonator the force measurement bandwidth = 35 kHz. For comparison, the authors of measurements using a low-Finesse cavity\cite{BartonNL2012} report a position sensitivity of $S_{N}^{1/2} = $ 6 $\times 10^{-13}$ m $\mathrm{Hz}^{-1/2}$ for an optically cooled resonator, where the measurement is intrinsically limited by photodetector noise.  

Referring to Fig.\ref{fig3}(b), measurements on a 30 $\mu m$ diameter resonator have sufficient SNR to resolve the $1^{st}, 2^{nd}$ and $3^{rd}$ mechanical resonances in the thermal noise driven motion with frequencies of $\omega_{m}/2\pi = $ 2.25, 3.36 and 3.95 MHz and $Q$ factors of 528, 884 and 927 respectively. The frequencies of these modes match reasonably well to previous measurements and the predictions of a simple model for the resonant frequency of circular drum resonances, expected at 1.59 and 2.14 times the fundamental frequency\cite{BartonNL2012}. The ability to resolve multiple mechanical modes is potentially useful for mass sensing\cite{HanayNATN2012} and for characterising the non-linear mechanical properties of the resonators\cite{Westra2010}. 
 
\begin{figure}[t]
  \begin{center}
  \includegraphics[width=\columnwidth]{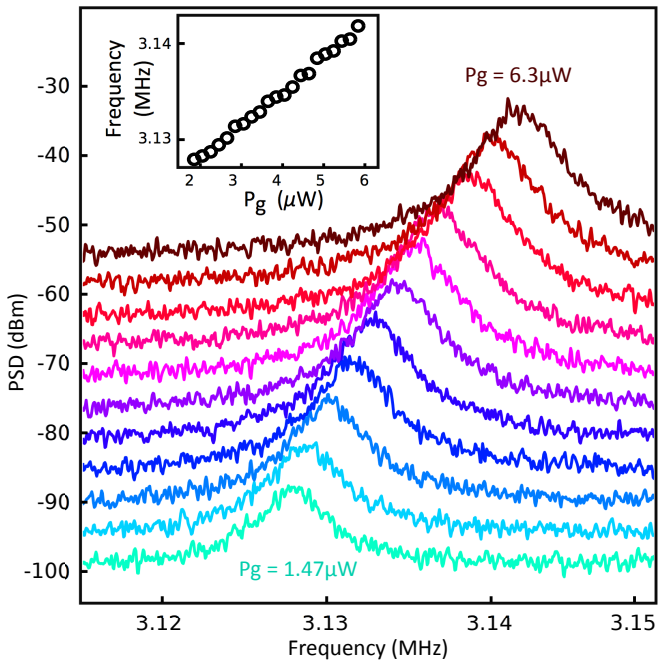}
\end{center}
\caption{\label{fig4} Mechanical resonance spectra of a 30 $\mu m$ diameter resonator with optical power dissipated by the graphene $\mathrm{P_{g}}$, with (inset) extracted resonance peak $\omega_{\mathrm{eff}}$. Spectra offset by 2.5 dBm on the vertical scale. }
\end{figure}

For applications it is desirable to be able to tune the mechanical frequency of the graphene resonator. Here we demonstrate this via the photo-thermal tension ($\sigma_{pth}$) induced in the membrane by optical absorption. Motion readout spectra in Fig.\ref{fig4} shows the tuning of the mechanical resonance $\omega_{m}$ as a function of optical power dissipated by the graphene $\mathrm{P_{g}}$. We define the effective mechanical frequency as $\omega_{\mathrm{eff}} \approx \omega_{m} \left( 1 +  \frac{\sigma_{pth}}{2 \sigma_{0}}   \right)$ which is valid in the limit $\sigma_{pth} \ll \sigma_{0}$, where $ \sigma_{0}$ is the intrinsic tension in the resonator membrane. According to theory\cite{BartonNL2012}, photons absorbed by the graphene induce a tension in the resonator membrane $\sigma_{pth} = A\mathrm{P_{g}}$ where $A$ is a material constant which depends on the absorption, thermal conductivity and thermal expansion coefficients of graphene. Fitting\cite{MonifiJLT2012} $\mathrm{P_{g}}/\mathrm{P1} = \left(\frac{4 \kappa_{ext} \kappa_{g}}{\kappa^{2} + 4\Delta^{2} }\right)$, yields $\mathrm{P_{g}}$ used in the inset plot of Fig.\ref{fig4}. From a linear fit to $\omega_{\mathrm{eff}}$ v.s $\mathrm{P_{g}}$ a value of $A$ = 9.3 N/(m W) is extracted, which is in reasonable agreement with a value of $A$ = 15 N/(m W) calculated using the fundamental constants of graphene\cite{BartonNL2012}. Furthermore $\mathrm{P_{g}}$ causes in increase in temperature T of the resonator $\bm{\mathit{\Delta}} T = \beta \mathrm{P_{g}}$ where the constant of proportionality $\beta = (2 \pi t k)^{-1}$ and $k$= 5000 W/(m K) is the thermal conductivity\cite{Geim2009} of graphene with $t$=0.335 nm the thickness of a monolayer. The maximum value of $\mathrm{P_{g}}= 6.3 \mu$W gives $\bm{\mathit{\Delta}} T$ = 0.59 K, which is a small change relative to room temperature (300 K). Therefore we conclude that optical heating will not significantly degrade the sensitivity $F_{min}$ of force measurements using this resonator. 
 
State of the art silicon cantilevers achieve a force sensitivity of $F_{min} = 7.4 \times 10^{-17}$ N ${\mathrm{Hz}}^{-1/2}$ at room temperature\cite{Gavartin2012}, owing to their high mechanical $Q$ factors. However, their relatively large mass places an intrinsic limit on their sensitivity. In contrast, improved high $Q$ factor graphene resonators would constitute ultra-sensitive force sensors with exceptionally small physical dimensions. A straightforward route to improving the optical readout (and hence the measurement bandwidth) is to optimise the $Q$ factor and optical mode volume of the microsphere, simultaneously reducing $\kappa$ and increasing $G$. Furthermore, by switching to homodyne detection an improvement in readout sensitivity is readily attainable. Finally, appropriate band-gap engineering of the graphene could further mitigate the negative effects of optical absorption. 

In conclusion, we have demonstrated the readout of graphene NEMS motion at room temperature using cavity enhanced evanescent sensing with high $Q$ optical microspheres. This approach enables ultra-sensitive readout of the graphene oscilliatory motion, and paves the way for high sensitivity/large bandwidth room temperature force measurements in a resonator mass regime not currently attainable using silicon NEMS devices. We present the first measurement of the optomechanical coupling coefficient $G$ of graphene at visible wavelengths, and show that material absorption does not result in poor readout sensitivity, and only minimally heats the resonator. Finally, we exploit the optical absorption to tune the mechanical frequency of the graphene resonator, and propose routes to improving the sensitivity of readout measurements.   

 \begin{acknowledgments}
This research was funded by the Australian Research Council Centre for Engineered Quantum Systems Grant No. CE110001013 and Discovery Project DP140100734. SEM imaging was performed at the Queensland node of the Australian National Fabrication Facility, a company established under the National Collaborative Research Infrastructure Strategy to provide nano and micro-fabrication facilities for Australia's researchers. Research at Cornell was supported by the NSF under DMR 1120296 and 1202991.
\end{acknowledgments}

% add bib
\bibliography{library.bib}
\bibliographystyle{apsrev}

\end{document}